\UseRawInputEncoding
\documentclass[
superscriptaddress,
preprint,
bibnotes,
amsmath,amssymb,
aps,
pra,
]{revtex4-1}
\usepackage{float}
\usepackage{graphicx}
\usepackage{epstopdf}
\usepackage{dcolumn}
\usepackage{bm,amsmath,amssymb}
\usepackage{latexsym,amsbsy}
\usepackage[colorlinks, linkcolor=blue, anchorcolor=blue, citecolor=blue, urlcolor=blue]{hyperref}

\usepackage{geometry}

\begin{document}

\title{Quantum battery of interacting spins with environmental noise}

\author{Fang Zhao}
\affiliation{Center for quantum technology research, School of Physics, Beijing Institute of Technology, Beijing 100081, People's Republic of China}
\affiliation{China Academy of Engineering Physics, Beijing 100088, People's Republic of China}

\author{Fu-Quan Dou}
\email[Corresponding author: ]{doufq@nwnu.edu.cn}
\affiliation{College of Physics and Electronic Engineering, Northwest Normal University - Lanzhou, 730070, China}

\author{Qing Zhao}
\email[Corresponding author: ]{qzhaoyuping@bit.edu.cn}
\affiliation{Center for quantum technology research, School of Physics, Beijing Institute of Technology, Beijing 100081, People's Republic of China}

\date{\today}

\begin{abstract}
A quantum battery is a temporary energy-storage system. We constructed the quantum battery model of an N-spin chain with nearest-neighbor hopping interaction and investigated the charging process of the quantum battery. We obtained the maximum energy in the quantum battery charged by a coherent cavity driving field or a thermal heat bath. We confirmed that for a finite-length spin chain, thermal charging results in a nonzero ergotropy, contradicting a previous result: that an incoherent heat source cannot charge a single-spin quantum battery. The nearest-neighbor hopping interaction induces energy band splitting, which enhances the energy storage and the ergotropy of the quantum battery. We found a critical point in the energy and ergotropy resulting from the ground-state quantum phase transition, after which the energy significantly enhance. Finally, we also found that disorder increased the energy of the quantum battery.
\end{abstract}

\maketitle

\section{Introduction}

%History[4-6]

~\\

A quantum battery (QB) system can potentially provide temporary energy storage.
The initially proposed QB was a two-level system that stores energy from an extra field \cite{origin}, and the initial physical model was proposed as a single two-level spin system.
Further QB studies verified that an N spin with an extra field increases the charging power of the QB \cite{ct0,compare}.
A primary goal of QB studies is finding the maximum energy stored in the QB during the charging time and increasing the energy release of the QB after the charging process \cite{ct0,ct1,ct2,ct3,ct4,ct5,ct6,ct7,ct8,ct9,ct10}.
Another research focus is the entanglement and work-extraction capability of the QB \cite{entang1,entang2,entang3,entang4,entang5,work1,work2,Dou_2020}.
As the energy released in a QB is usually in a thermal heat bath, a QB has commonly discussed the energy for work in terms of ergotropy \cite{ergotropy}.

In the usual case, an extra field charges the energy of the QB.
The energy exchange between the QB and a cavity field occurs by coupling of the QB with an energy-charged cavity field in an excited energy state \cite{ct0}.
Energy oscillations during the charging process necessitate accurate control of the charging time.
Other charging sources are directly charged by a magnetic field or a thermal heat bath \cite{ct4,ct8,magnetic2,open1}.
A QB charged in a thermal bath must be discussed in an open quantum system.
A single-atom QB can be energy charging by a thermal heat bath, but its useful energy for work is always zero \cite{open1}.
Due to the decay rate and the driving field, the energy-charging process of the QB gradually stabilizes in an open quantum system \cite{open1,open2,int1,open3,entang5,open5,open6,open7,open8}, so controlling the charging time is not necessary.

Few previous studies have considered the hopping interaction between each spin in a QB \cite{int1,int2,int3,int4}.
In studies that do consider such interactions, the energy charging and release by the QB are not discussed.
In a real spin chain model, the hopping interaction is vital and cannot be ignored.
The spin chain interaction creates a ground-state quantum phase transition and influences the ground-state properties of the QB.
The quantum phase transition also influences the energy charging and release of the QB.
The most simple interaction in a spin chain is the nearest hopping interaction.

\begin{figure}
\centering
  \scalebox{0.33}{\includegraphics{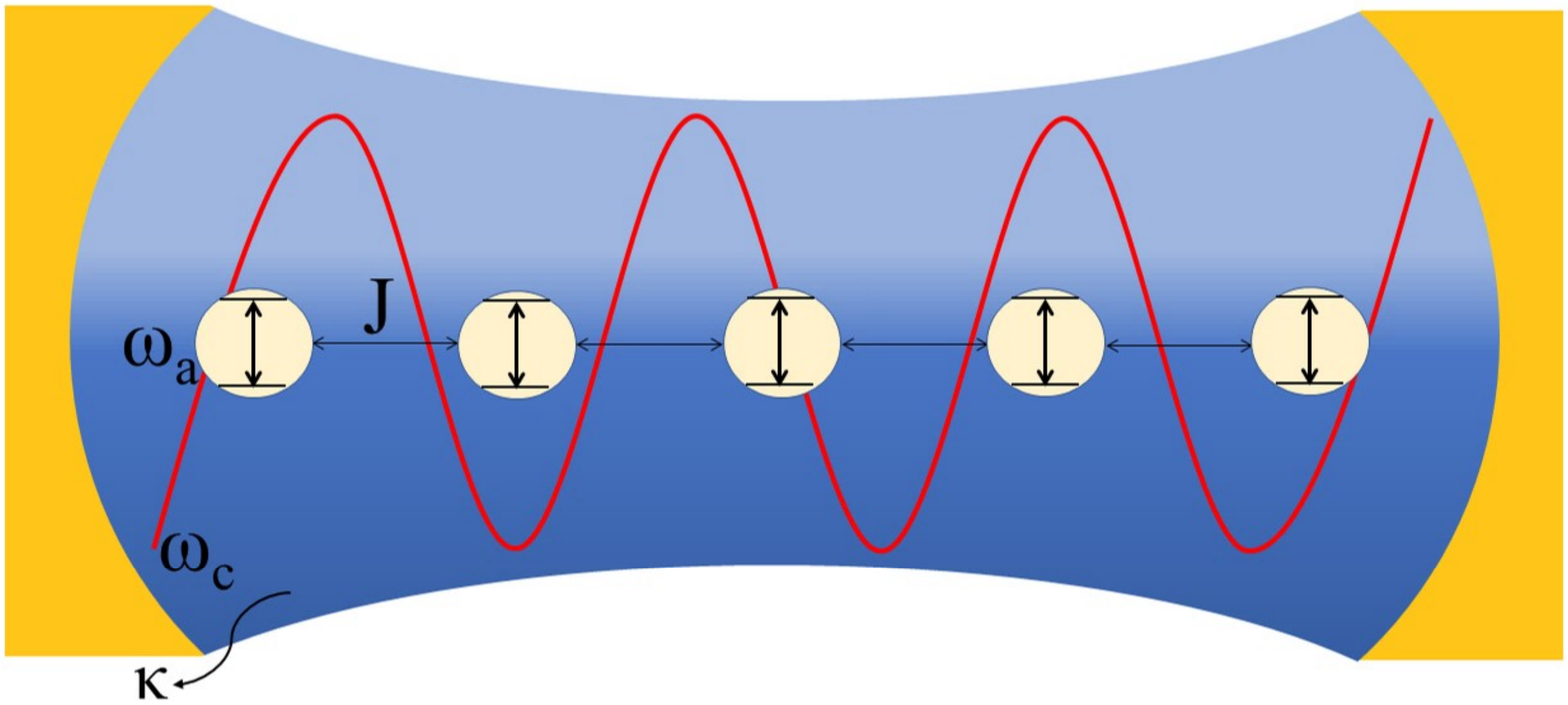}}
  \caption{
A schematic diagram of the QB charging protocol used in this study.
It includes an N spin with a frequency of $\omega_a$.
The spin has a nearest-neighbor hopping interaction with a strength of $J$.
The spin is coupled with a single-photon cavity with a frequency of $\omega_c$ and a decay rate of $\kappa$.
}\label{fig.model}
\end{figure}

In this paper, we discuss a QB undergoing nearest-neighbor hopping interaction, and the quantum-phase influence of the QB with a hopping term.
Fig.\ref{fig.model} shows the charging protocol of our QB system.
The physical model of the QB is an $N$-spins Dicke-spin chain with hopping interaction $J$.
The QB is coupled to a cavity field with the decay rate of the $\kappa$-boson heat bath.
Because $\kappa$ decays at a certain rate, we must investigate the charging of the QB in an open quantum system.
The energy charging of the QB is accomplished by an extra coherent driving field or is directly provided by the thermal heat bath.
We study the energy and ergotropy of the QB with and without the nearest-neighbor hopping interaction.
We compare the difference between the two types of charging protocol: charging from an extra coherently driven field and directly charging from a thermal heat bath.
We investigate how the hopping interaction and the spin chain length influences the energy and ergotropy of the QB.
Finally, we also study how disorder influences the performance of the QB.

This paper organized as follows.
In Sec. \ref{sec.model}, we introduce the QB model and the dynamic equations.
We define the energy and ergotropy of the QB.
In Sec. \ref{sec.charging}, we report on the dynamics of QB charging without and with the hopping interaction, respectively.
We also investigates how the ground-state quantum phase transition (QPT) influences the QB.
In Sec. \ref{sec.disorder}, we discuss the QPT influence on the charging time of the QB, and the disorder influence on the energy and ergotropy of the QB.

\section{Model}\label{sec.model}

The QB is modeled by an $N$-spin chain, as shown in Fig. \ref{fig.model}.
The hopping interaction strength between nearest-neighbor spins is $J$.
The spin is embedded in a microcavity with a cavity loss rate of $\kappa$.
The Rabi frequency of the photon–spin interaction is $g$.
The total Hamiltonian of this QB system is:
\begin{align}
  H_S&=H_A+H_B+H_I,\\
  H_A&=\omega_c c^\dag c,\\
  H_B&=\omega_a \sum_{i=1}^{N} \sigma_+^i\sigma_-^i+J\sum_{i=1}^{N-1} (\sigma_+^i\sigma_-^{i+1}+h.c.),\\
  H_I&=\sum_{i=1}^{N} g(\sigma_+^{i}c+h.c.).
\end{align}
In the above expressions, $H_A$ is the Hamiltonian of the cavity part with annihilation (creation) operator $c (c^\dag)$.
$\omega_c$ is the cavity frequency of the cavity field.
$H_B$ is the Hamiltonian of the QB with $\sigma_\pm^{i}$ as the raising/lowering spin operator for the $i$-th spin and an atom frequency of $\omega_a$.
$J$ is the nearest hopping interaction.
$H_I$ is the interaction term between the spin and the cavity field with spin–photon coupling constant $g$.

Within this set-up, the cavity field is driven by an extra classical field that charges the QB.
The energy input from the driving field later transfers to the QB via the cavity–spin interaction.
In this paper, the QB is charged in two ways: first from an extra coherent driving field acting on the cavity, and second from a thermal heat bath coupled with the cavity field.
We denote these two charging approaches as coherent and thermal charging, respectively.

The Hamiltonian of the coherent driven field is
\begin{equation}
  H_d'=f(e^{-i\omega_dt}c^\dag +e^{i\omega_dt}c).
\end{equation}
Here, $f$ is the driving field strength.
After a unitary transformation $U=e^{i\omega_ct}$, the explicit time dependence term can be removed into \cite{open1}:
\begin{equation}
  H_d'=f(e^{-i\delta t}c^\dag +e^{i\delta t}c),
\end{equation}
where $\delta=\omega_d-\omega_c$.
When we take $\omega_c=\omega_d$ for simplicity, the driven field Hamiltonian will reduce to
  \begin{equation}
  H_d=f(c^\dag +c).
\end{equation}
The dynamic process of the QB coherent charging is obtained by solving the Lindblad master equation:
\begin{equation}\label{eq.coherent_master}
  \dot{\rho}_S(t)=-i[H+H_d,\rho_S(t)]+\kappa L_{c}[\rho_S],
\end{equation}
where $\kappa$ is the decay rate of the cavity field, and $L_c[\rho_S]=c\rho_S c^\dag-\frac{1}{2}(c^\dag c\rho_S+\rho_Sc^\dag c)$ is the Lindblad super-operator.

The dynamics of thermal charging are obtained by solving the Lindblad master equation:
\begin{equation}\label{eq.thermal_master}
  \dot{\rho}_S(t)=-i[H,\rho_S(t)]+\kappa(n_B+1) L_{c}[\rho_S]+\kappa n_BL_{c^\dag}[\rho_S].
\end{equation}
Where $n_B=1/(\exp[\omega_c/(k_BT)]-1)$ is the mean occupation number of the boson heat bath.

The charging process of the QB fills the empty QB from the cavity field.
We prepare an empty QB by initializing the spin in its ground-state $|g\rangle_B$.
The initial state of the cavity is the vacuum state $|0\rangle_A$. Thus, the initial state of the whole system is:
\begin{equation}
  |\psi(0)\rangle=|0\rangle_A\otimes|g\rangle_B,
\end{equation}
where $|\psi(0)\rangle$ is the initial state of the whole system, and $|g\rangle_B$ is the initial state of the QB corresponding to the energy ground-state of $H_B$.
When the cavity two-level interaction $g$ is turned on, the charging process immediately starts the energy exchange between the spin chain and the cavity field.

%\section{Ergotropy and Efficiency}\label{sec.energy}

The energy storage in the QB at time $t$ is given by:
\begin{equation}\label{eq.EB}
  E_B(t)=\text{tr}[H_B\rho_B(t)],
\end{equation}

where $\rho_B(t)=\text{tr}_A[\rho_{S}(t)]$ is the reduced density matrix of the QB at time $t$.
The energy-charged into the QB is $E_B(t)-E_B(0)$, where $E_B(0)=E_G$ is the ground-state energy of the QB.
Therefore, the actual charging energy of the QB is
\begin{equation}\label{eq.deltaE}
  \Delta E(t)=E_B(t)-E_G.
\end{equation}

\begin{figure*}
\centering
  \scalebox{0.61}[0.61]{\includegraphics{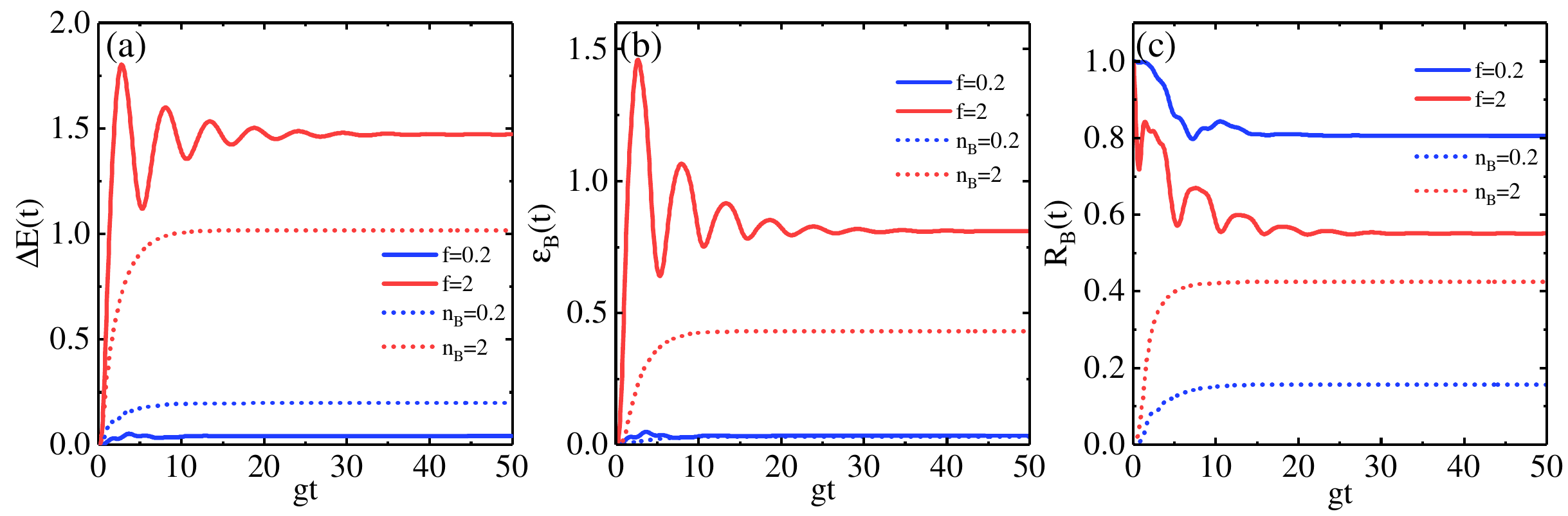}}
  \caption{
  The dynamic charging process of the QB.
  (a) Energy $\Delta E(t)$ (b) Ergotropy $\varepsilon_B(t)$ and (c) Efficiency $R_B(t)$ as the function of $gt$.
  Other parameters are $N=3$, $\omega_a=\omega_c=g=\kappa=1$ and $J=0$.
    }\label{fig.evolution}
\end{figure*}

The charging energy $\Delta E(t)$ can characterize the property of the QB.
However, under the second law of thermodynamics, it cannot be transformed into work without dissipating the heat.
Ergotropy characterizes the ability of the QB to generate useful work \cite{ergotropy}.
The ergotropy is defined as
\begin{equation}\label{eq.ergotropy}
\varepsilon_B(t)=E_B(t)-\min_U \text{tr}[H_BU\rho_B(t) U^\dag].
\end{equation}
Diagonalizing $H_B$ and $\rho_B(t)$, we respectively obtain
\begin{align}
  \rho_B(t)&=\sum_n r_n(t) |r_n(t)\rangle \langle r_n(t)|, \\
  H_B&=\sum_n e_n |e_n\rangle \langle e_n|.
\end{align}

The eigenvalues of $\rho_B(t)$ are arranged in descending order as $r_0\geq r_1\geq \cdots $, and the eigenvalues of $H_B$ are arranged in ascending order as $e_0\leq e_1 \leq\cdots$.
The term $\min_U \text{tr}[H_BU\rho_B(t) U^\dag]$ of Eq.(\ref{eq.ergotropy}) can be simplified as follows \cite{ergotropy,open1}:
\begin{equation}\label{eq.min}
  \min_U \text{tr}[HU\rho(t) U^\dag]=\sum_n r_n e_n.
\end{equation}

It is easily proved that $\varepsilon_B(t)$ is always non-negative and smaller than $\Delta E(t)$.
Therefore, we can define the following efficiency $R_B(t)$ as the percentage of $\varepsilon_B(t)$ among the total charging energy $\Delta E(t)$:
\begin{equation}
  R_B(t)=\frac{\varepsilon_B(t)}{\Delta E(t)}.
\end{equation}

Using the energy $\Delta E$, we can judge the charging energy of the QB.
The ergotropy could judge $\varepsilon_b$ is the useful energy-releasing for useful work of the QB.
Furthermore, the efficiency $R_B$ describes the useful energy release efficiency of the QB.
This paper focuses on $\Delta E$, $\varepsilon_b$, and $R_B$ during the QB charging process.

\section{The Charging property of the QB}\label{sec.charging}

This section will discuss the charging properties of the QB with or without the hopping interaction.
We will first discuss QB's dynamic and find the maximum energy and the ergotropy during the charging process.
Then discuss the quantum phase transition induced by the hopping interaction.
Finally, We also discuss how the quantum phase influences the charging properties of the QB.

\subsection{Charging Without Hopping Interaction}

Before discussing the effect of the hopping interaction, we discuss the charging properties of the QB without the hopping term $J=0$.
The ground-state of the QB in this case is
\begin{equation}
  |g\rangle_B=|0\rangle^{\otimes N},
\end{equation}
where $|0\rangle$ is the ground-state of a single spin. The ground-state energy $E_G$ of the QB without the hopping interaction is zero.

\begin{figure*}
\centering
  \scalebox{0.61}[0.61]{\includegraphics{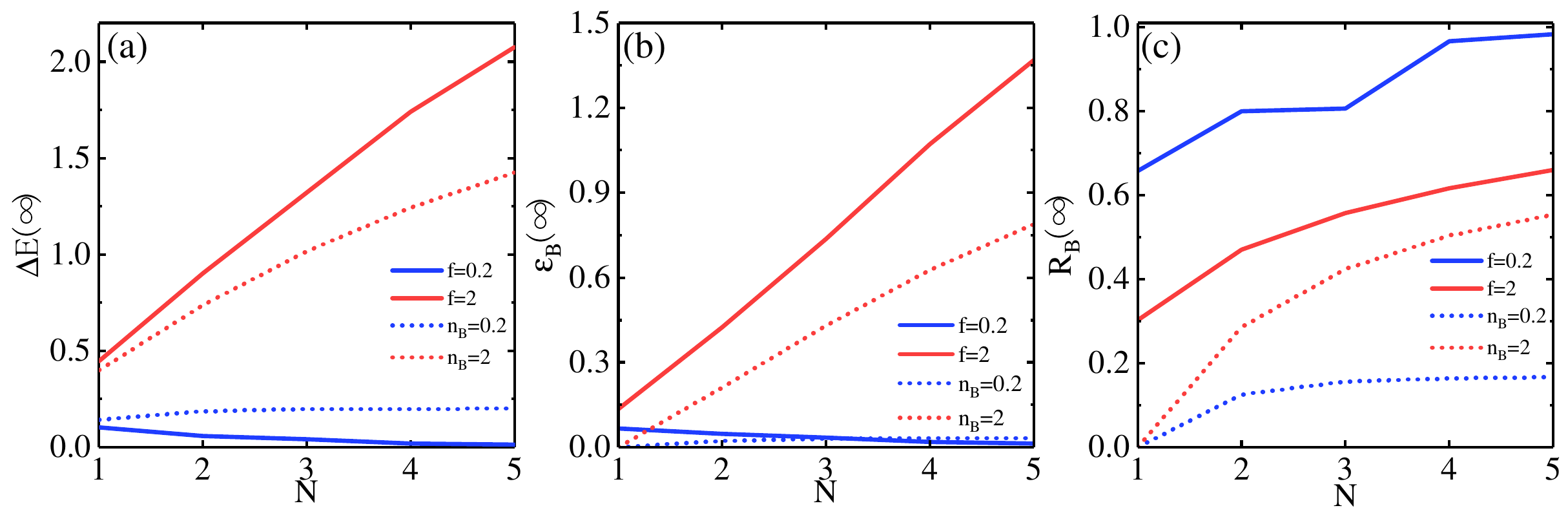}}
    \caption{
  The QB steady-state energy, ergotropy, and the energy release rate for the different number of spins.
  (a) QB energy $\Delta E(\infty)$ (b) Ergotropy $\varepsilon_B(\infty)$ (c) Efficiency $R_B(\infty)$.
  Other parameters are $\omega_a=\omega_c=g=\kappa=1$ and $J=0$.
    }\label{fig.number}
\end{figure*}

The dynamics of the coherent and thermal charging processes are determined by Eqs. (\ref{eq.coherent_master}) and (\ref{eq.thermal_master}), respectively (see Fig. \ref{fig.evolution}).
$\omega_a, \omega_c, \kappa$, and $g$ are not relatively essential parameters for the system stable-state.
Therefore in this paper, we only discuss the most direct case, where
$\omega_a=\omega_c=g=\kappa=1$.
Owing to the decay loss, the strong and weak driving strengths differ during the charging process.
To compare these two strengths, we set $f=2$ ($n_B=2$) and $f=0.2$ ($n_B=0.2$).
The strong and weak are just is for compared with $\kappa$.
As shown in panels (a) and (b) of Fig. \ref{fig.evolution}, the energy $\Delta E(t)$ and ergotropy $\varepsilon_B(t)$ of the QB were gradually stabilized.
Under both coherent and thermal charging, the energy and ergotropy of the QB increased with driving strength, but the coherent charging oscillated in the beginning phase.
The steady-state energy at infinite time was the maximum charging energy.
However, the ergotropy of the QB was maximized within the oscillatory period, indicating that the oscillations boosted the efficiency of the coherently charged QB at the beginning.
In contrast, both the energy and ergotropy of the thermally charged QB were maximized at steady-state after infinite time.
Besides, the QB efficiency in the thermal charging large driving strength will also correspond to a significant efficiency opposite to the thermal charging.

We could also find a nonzero ergotropy during the charging process for the thermal charging.
This result differs from those of previous works on simple-spin systems, in which the ergotropy $\varepsilon_B$ is always zero \cite{open1}.
A nonzero ergotropy can be obtained by multiplying the spin of the QB with the thermal charging, as shown in Fig. \ref{fig.number}.
We verified the thermal charging of a single spin ($N=1$) QB always induces a zero ergotropy, as reported in previous research \cite{open1}.
However, thermal charging of a finite-length chain ($N\geq2$) QB induces a nonzero ergotropy.
Under strong charging conditions, the spin numbers increase the energy, ergotropy, and efficiency of the QB, but under weak charging conditions, the energy and ergotropy of the QB both decreased with increasing chain length.

\begin{figure}
\centering
  \centering
  \includegraphics[width=12cm]{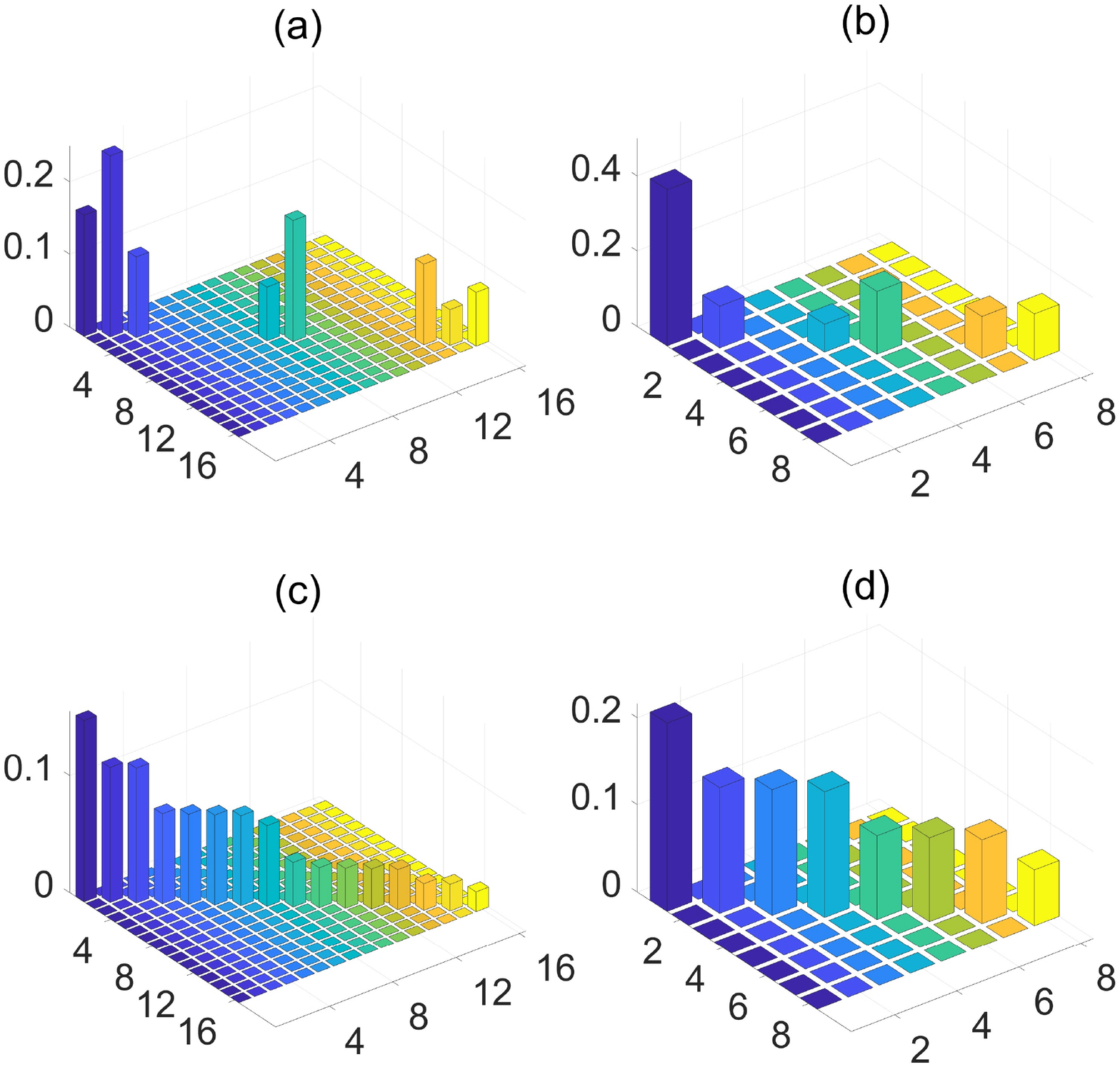}
  \caption{
  Steady-state density matrices in the energy representation: Panels (a) and (b) show the steady-state densities of the whole system and the reduced density matrix of the QB under thermal charging, respectively. Panels (c) and (d) show the thermal state density matrices of the whole system and the QB, respectively. Other parameters are $N=3$, $\omega_a=\omega_c=g=\kappa=1$, $J=0$ and $n_B=2$.
    }\label{fig.thermal}
\end{figure}

\begin{figure*}
\centering
  \scalebox{0.61}[0.61]{\includegraphics{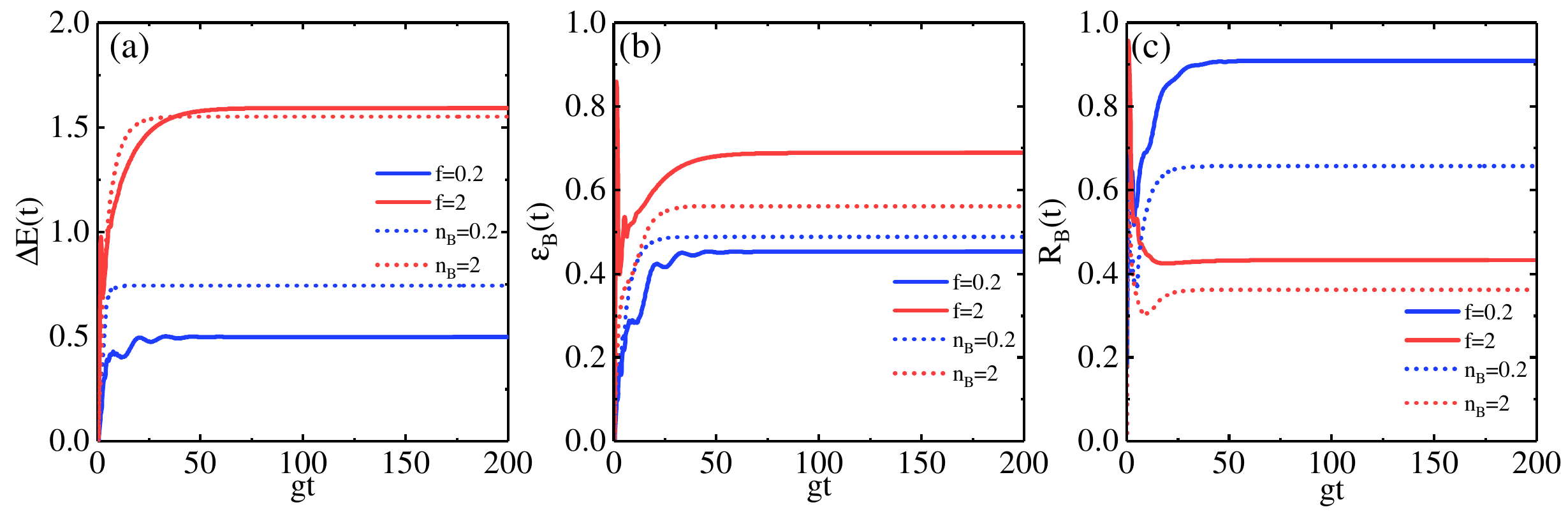}}
  \caption{
  The QB dynamic charging process with considering the hopping interaction.
  (a) Energy $\Delta E(t)$ (b) Ergotropy $\varepsilon_B(t)$ and (c) Efficiency $R_B$ as the function of $gt$.
  Other parameters are $N=3$, $\omega_a=\omega_c=g=\kappa=1$ and $J=1$.
    }\label{fig.evolution_J}
\end{figure*}

Above, we explained that thermal charging can induce nonzero ergotropy.
If the QB exists in the thermal state $\rho_B=\rho_B^{th}=\frac{e^{-\beta H_B}}{z}$, where $z=tr[ e^{-\beta H_B}]$, it is easily verified that the ergotropy is always zero.
The thermal-state density matrix $\rho_B^{th}$ and the Hamiltonian representation of the energy are respectively given by
\begin{align}
  \rho_B^{th}&=\frac{1}{z}\left(
           \begin{array}{cccc}
             e^{-\beta e_1} & 0 & \cdots&0 \\
             0 & e^{-\beta e_1} & \cdots&0 \\
             0 & 0 & \cdots& e^{-\beta e_n}\\
           \end{array}
         \right), \nonumber \\
         ~~~
  H_B&=\left(
           \begin{array}{cccc}
             e_1 & 0 & \cdots&0 \\
             0 & e_2 & \cdots&0 \\
             0 & 0 & \cdots& e_n\\
           \end{array}
         \right).
\end{align}
Here, the energy eigenvalues are in ascending order, and the diagonal elements of the thermal state density matrix are in descending order.
Thus, the mean thermal- state energy is $E_B=\text{tr}[H_B\rho^{th}]=\sum_n r_n e_n$ (equaling the right hand of Eq.(\ref{eq.ergotropy})), meaning that the ergotropy of a thermal state is always zero.
When the ergotropy is nonzero, the density of the QB is reduced and the thermal state is not attained.
In this scenario (Fig. \ref{fig.thermal}), the steady-state density matrix $\rho_B$ and system density matrix $\rho$ of the QB differ from their corresponding density matrices $\rho_B^{th}$ and $\rho^{th}$ in the thermal state.
As thermal driving is coupled only to photons within a $\beta$ temperature thermal bath, the total photon-spin system cannot be purified into the thermal state.
The reduced density matrix $\rho_B$ also cannot exist in the thermal state. Thus, the ergotropy of a non-single-atom QB can be nonzero under thermal charging conditions.

\begin{figure*}
\centering
  \scalebox{0.61}[0.61]{\includegraphics{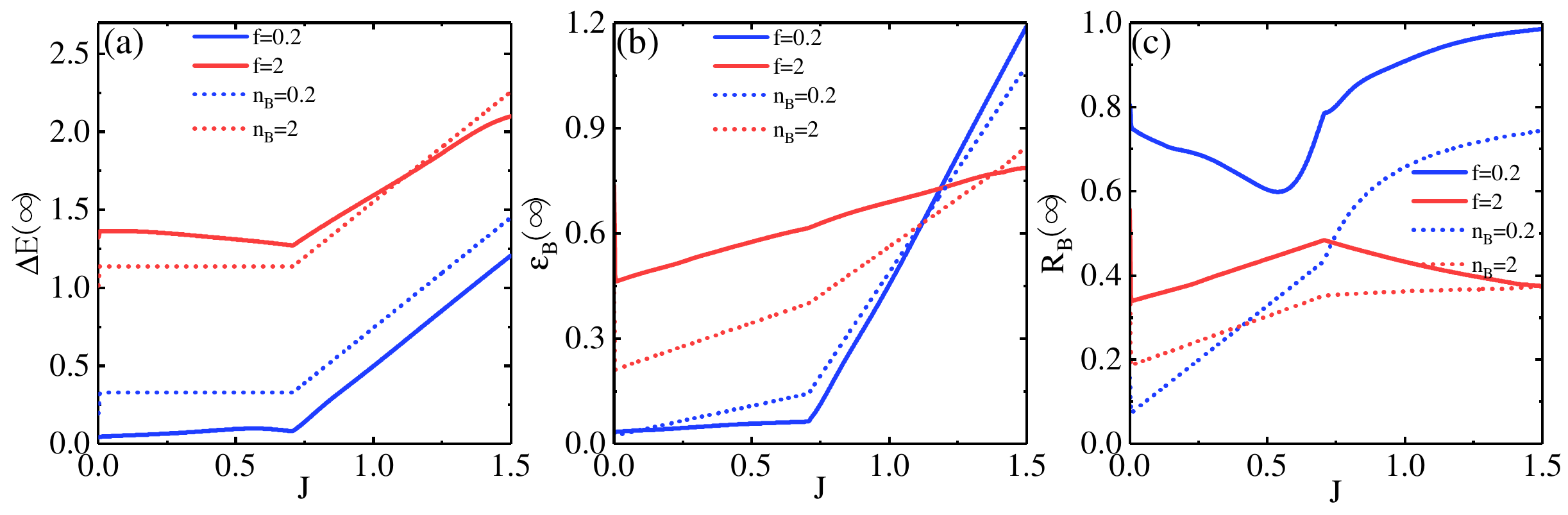}}
\caption{
  The QB steady energy and ergotropy for different hopping interaction.
  (a) Energy $\Delta E(\infty)$ (b) Ergotropy $\varepsilon_B(\infty)$ and (c) Efficiency $R_B(\infty)$ as a function of $J$.
  Other parameters are $N=3$, $\omega_a=\omega_c=g=\kappa=1$.
  }\label{fig.interaction}
\end{figure*}

\subsection{Charging with Hopping Interaction}

In the previous subsection,
we discussed the coherent and thermal charging processes of the QB without the hopping interaction.
We here consider a spin-chain QB with nearest-neighbor interacting strength $J$ in the QB.
The initial state of the QB determines the energy ground-state of the interacting spin chain.
Through investigating the dynamics of the coherent and thermal charging processes, we discuss how the hopping interaction influences the energy and ergotropy of the QB.

When considering the hopping interaction, the charging dynamics of the QB are shown in Fig. \ref{fig.evolution_J}.
The charging dynamics of the QB were largely unaffected by hopping interaction, but the hopping interaction increased the time at which the QB received its maximum energy, and also the energy of the QB under coherent and thermal charging conditions.
However, the hopping interaction decreased the ergotropy of the QB.
Weak charging reduced the efficiency of both coherent and thermal charging. Meanwhile, the hopping interaction did not remove the oscillations from the coherent charging phase.
Regardless of whether considering the hopping interaction or not, the difference between coherent and thermal charging is only the energy oscillations during the charging process.

\begin{figure*}
\centering
  \scalebox{0.66}[0.66]{\includegraphics{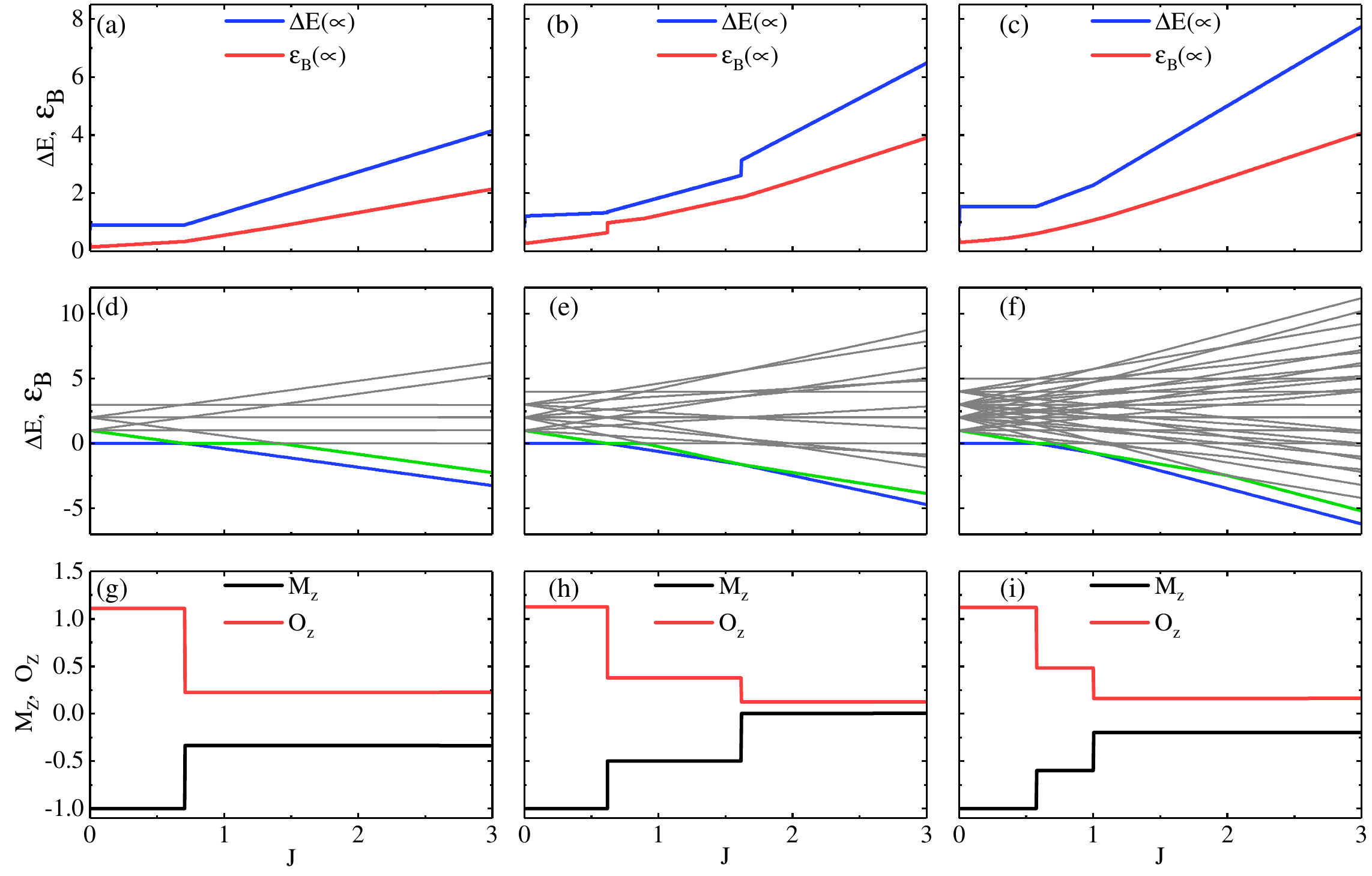}}
  \caption{
  (a), (b), and (c) are the QB energy $\Delta E(\infty)$ and ergotropy $\varepsilon_B(\infty)$ as a function of the hopping interaction for the spin number N = 3, 4, 5, respectively.
  (d), (e), and (f) are the respective energy spectra.
  (g), (h), and (i) are the corresponding ordering parameters.
  The other parameters are $n_B=1$, $g=\kappa=1$.
}\label{fig.spectrum}
\end{figure*}

To further discuss the influence of the hopping interaction, we calculated the steady-state energy and the ergotropy of the QB at different hopping interaction as shown in Fig. \ref{fig.interaction}.
At $J=0$, the energy and ergotropy were remarkably changed by the interaction term, which broke the symmetry of the QB.
A non-differentiable points appeared at $J=1/\sqrt{2}$.
Before this point, the QB energy was a nearly constant function of hopping interaction, but thereafter, it significantly increased with hopping interaction.
Meanwhile, the ergotropy of the QB exhibited two non-differentiable points: one at $J=1/\sqrt{2}$, the other at $J=\sqrt{2}$.

We then calculated the steady-state energy and ergotropy of a thermally charged QB with different spin numbers ($N=3,4,$, and $5$) are shown in panels (a), (b), and (c), respectively, of Fig. \ref{fig.spectrum}.
The energy versus hopping interaction of the QB developed an extra non-differentiable point at large spin numbers, but the energy and ergotropy of the QB exhibited similar growth trends at small and large spin numbers.
Whereas the energy did not significantly increase before the first non-differentiable point, the ergotropy was a continuously increasing function of hopping interaction.

We calculated the energy spectra to further discuss the non-differentiable points in the QB's energy and ergotropy.
The results for $N=3,4$,and $5$ are shown in panels (d), (e) and (f) of Fig. \ref{fig.spectrum}, respectively.
Each non-differentiable point of the energy corresponded to a crossing of the ground-energy state.
Before the first ground-state energy crossing, the ground-state energy was always zero.
After first crossing point, the ground-state energy was gradually decreased by splitting of the energy bands.
From Eq.(\ref{eq.deltaE}), we found that the decreasing ground-state energy due to band splitting increased the energy by increasing the hopping interaction

The different energy behaviors of the QB's energy around the first non-differentiable point induced a QPT.
To further discuss the QPT, we introduce the order parameter.
One common order parameter is the mean magnetic field $M_z$, given by
\begin{equation}
M_z=\frac{\langle S_z \rangle_g}{N}.
\end{equation}
Here we define another order parameter $\xi_z$ as
\begin{equation}
  \xi_z=\frac{\langle S_z^2 \rangle_g}{N^2},
\end{equation}
where  $\langle\cdots\rangle_g$ represents the average on the ground-state, and the total spin operator is $S_z=\sum_{i=1}^{N} \sigma_z^{i}$.
We consider the only the ground-state’s ordering parameter because both the non-differentiable points of the energy and ergotropy correspond to the ground-state energy crossing.

The calculated order parameters for $N=3,4,$ and $5$ are presented in panels (g), (h) and (i) of Fig. \ref{fig.spectrum}, respectively.
The ordering parameter was discontinuous at the point $J=1/\sqrt{N}$ and $J=\sqrt{N}$.
Around these discontinuous points, indicating a first-order QPT at this point.
We have already found significant changes in the QB charging properties after the first non-differentiable point.
Before the first non-differentiable point, the order parameter was $M_z=-1$, meaning that each spin was in the spin-down state corresponding to the ferromagnetic phase.
After the first non-differentiable point, the ordering parameter $M_z>-1$ departed from the ferromagnetic phase.
The ground-state quantum phase significantly influenced the charging of the QB. In a non-interacting spin-chain QB, the ground-state of the QB was always spin-down.
Accordingly, the QB existed in the ferromagnetic phase, which is unsuitable for more energy storage.

\section{Charging time and disorder}\label{sec.disorder}

In the previous section, we discussed how hopping interaction influences the energy and ergotropy of the QB.
We also observe that when the hopping interaction was nonzero, the energy of the QB was enhanced after the first QPT point, and the charging speed of the QB significantly increased.
This section investigates the influence of hopping interaction on the charging speed, and the effects of onsite energy disorder

\begin{figure}
\centering
  \scalebox{0.32}[0.32]{\includegraphics{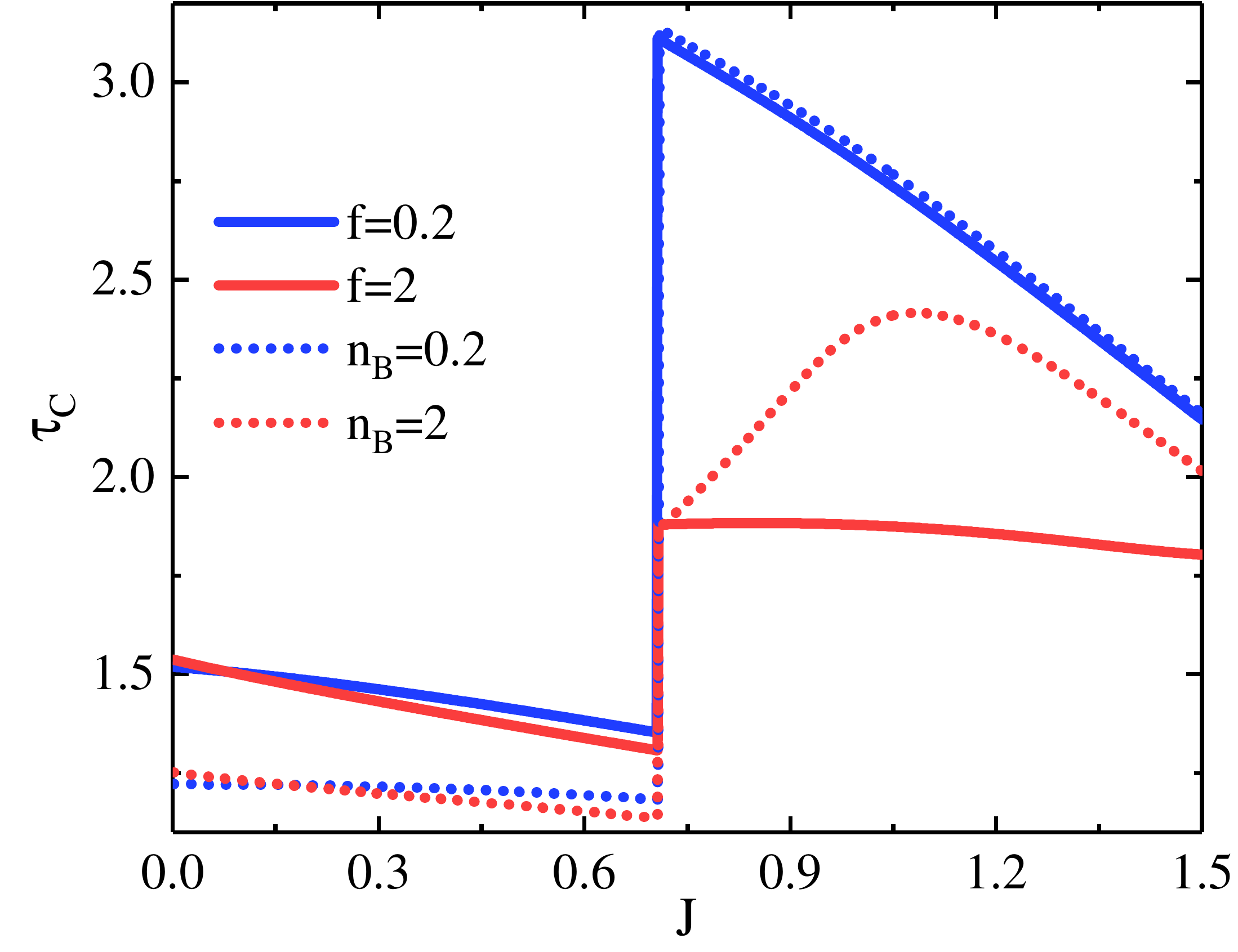}}
  \caption{
  The charging time $\tau_c$ as a function of the hopping interaction $J$.
  Other parameters are $N=3$, $\omega_a=\omega_c=g=\kappa=1$ .
}\label{fig.time}
\end{figure}

We here focus on the steady-state energy and ergotropy of the QB during charging in an open quantum system. Theoretically, the steady-state is reached only after infinite time.
In reality, the charging process is ceased when the energy and ergotropy of the QB are sufficiently close to their steady-state values.
To judge the charging speed, we must therefore define a charging time.
We first define the charging power as
\begin{equation}
  P_B(t) = \frac{\Delta E(t)}{t}.
\end{equation}
The charging time $\tau_c$ then defines the time at which the charging power is maximized:
\begin{equation}
\tau_c=\arg\max_t P_B(t).
\end{equation}

\begin{figure}
\centering
  \scalebox{0.38}[0.38]{\includegraphics{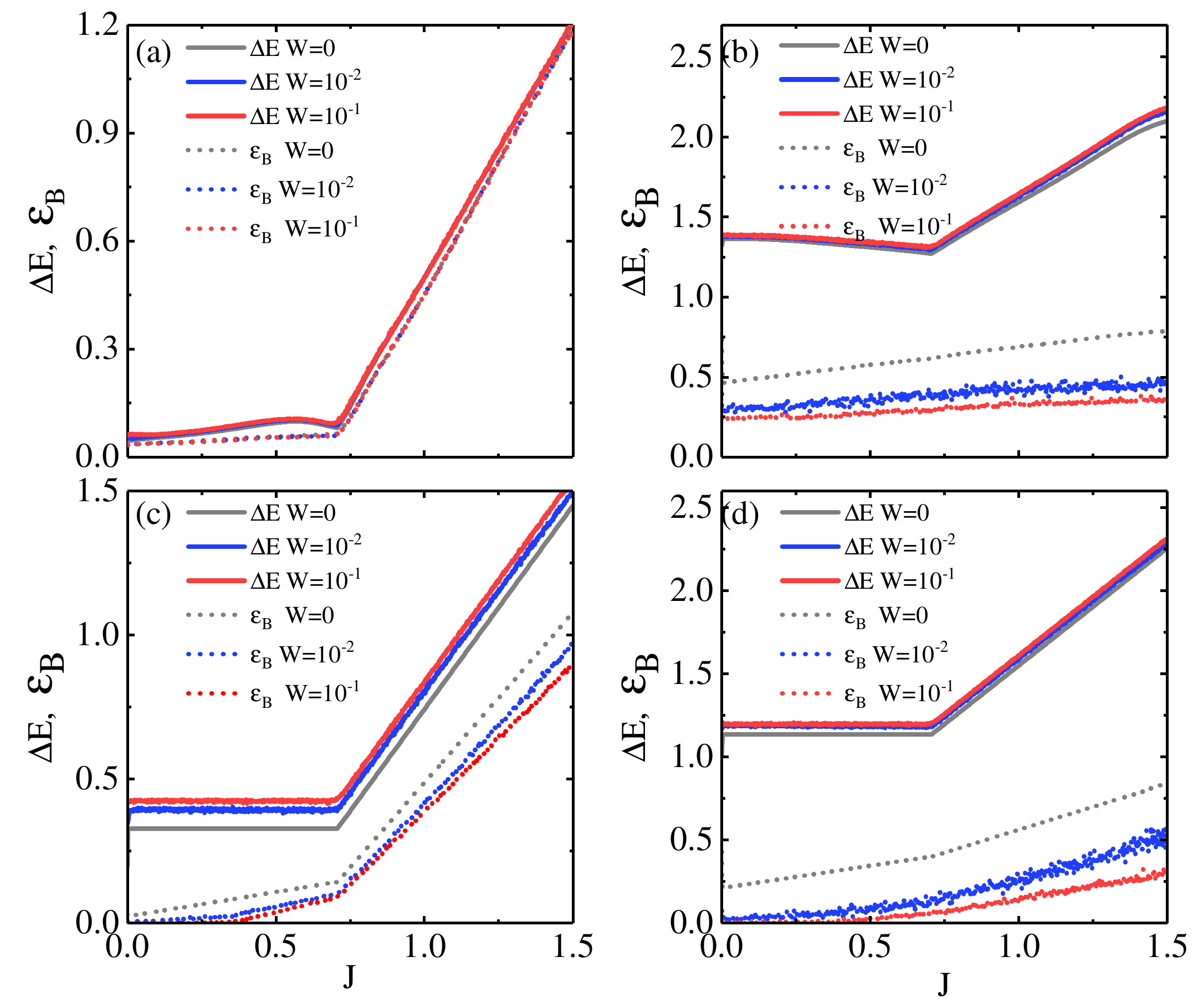}}
  \caption{
  The steady-state energy and ergotropy of the QB influenced by the onsite disorder $W$.
  These four figures correspond with different charging strengths with (a) $f=0.2$ (b) $f=2$ (c) $n_B=0.2$ (d) $n_B=2$.
  Other parameters are $N=3$, $\omega_a=\omega_c=g=\kappa=1$.
  }\label{fig.disorder}
\end{figure}

Fig. \ref{fig.time} shows the charging times $\tau_c$ at different hopping interaction.
We observe that the charging time $\tau_c$ suddenly increased after the phase transition point.
Before the phase transition point, the charging time gradually reduced with increasing hopping interaction.
After the QPT point, energy and ergotropy increased at the expense of increasing the charging time.

We now discuss the influence of disorder on the energy and ergotropy of the QB.
Here we consider only an onsite disorder in the free energy term, which changes the Hamiltonian of the QB as follows:
\begin{equation}
  H_B=\omega_a \sum_{i=1}^{N}(1+\delta_i) \sigma_+^{i}\sigma_-^{i}+J\sum_{i=1}^{N-1}(\sigma_+^{i}\sigma_-^{i+1}+h.c.),
\end{equation}
where $\delta_i\in[-W/2,W/2]$ is the onsite disorder and $W$ is the disorder strength.
To ensure accurate numerical calculations, the result was computed $100$ times and averaged to give the result in Fig. \ref{fig.disorder}.
We can find that the disorders do not affect the location of the QPT point.
After the phase transition point, the energy and ergotropy of the QB still increased with increasing hopping interaction, as observed previously.
The onsite disorder little affected the stability of the energy.
However, the ergotropy is unstable in the onsite disorder.
The increased disorder strength enhanced the energy of the QB.
The increased QB energy was similar in magnitude to the energy transfer enhanced by the disorder \cite{noise1,noise2,noise3}, although a large disorder would reduce the ergotropy.

\section{Conclusion}

We investigated QB charging in an open quantum system with coherent and thermal charging.
Without hopping interaction, we found that for spin lengths $N\geq2$, thermal charging imparted a nonzero ergotropy to the QB.
Under weak and strong charging conditions, increasing the spin length decreased and increased the ergotropy of the QB, respectively.
In the system with hopping interaction, the QB energy increased with charging time after the first ground-state energy crossing point.
The ground-state QPT also affected the energy of the QB.
Finally, we investigated the onsite disorders of QB, and verified that disorder increased the energy but decreased the ergotropy of the QB.

\begin{acknowledgments}
This work is supported by the National Science Foundation of China (Grants No. 11675014, No. 12075193).
\end{acknowledgments}

\bibliography{QB_manuscript}
\end{document}